\documentclass[%
 twocolumn,
 reprint,
showpacs,
 amsmath,amssymb,
 aps,
]{revtex4}

\usepackage{color}
\usepackage{ulem}
\usepackage{graphicx}
\usepackage{dcolumn}
\usepackage{bm}
\usepackage{here}
\usepackage{url}

\definecolor{myorange}{rgb}{0.7,0.5,0.0}
\definecolor{mygreen}{rgb}{0.0,0.7,0.0}

\newcommand{\beginsupplement}{%
        \setcounter{table}{0}
        \renewcommand{\thetable}{S\arabic{table}}%
        \setcounter{figure}{0}
        \renewcommand{\thefigure}{S\arabic{figure}}%
     }

\begin{document}

\preprint{APS/123-QED\author{Takahiro Ohgoe}}

\title{Competition among Superconducting, Antiferromagnetic, and Charge Orders \\ with Intervention by Phase Separation in the 2D Holstein-Hubbard Model}

\author{Takahiro Ohgoe}
\author{Masatoshi Imada}%
\affiliation{%
 	Department of Applied Physics, University of Tokyo, 7-3-1 Hongo, Bunkyo-ku, Tokyo 113-0033, Japan
}%

\date{\today}

\begin{abstract}
  Using a variational Monte Carlo method, we study competitions of strong electron-electron and electron-phonon interactions in the ground state of Holstein-Hubbard model on a square lattice. At half filling, an extended intermediate metallic or weakly superconducting (SC) phase emerges, sandwiched by antiferromagnetic (AF) and charge order (CO) insulating phases. By the carrier doping into the CO insulator, the SC order dramatically increases for strong electron-phonon couplings, but largely hampered by wide phase separation (PS) regions. Superconductivity is optimized at the border to the PS.
\end{abstract}

\pacs{63.20.Kr, 71.10.Fd, 71.27.+a, 74.25.Kc, 74.72.-h}
\maketitle


{\it Introduction}. ---The electron-phonon interaction in condensed matter is the origin of many important phenomena such as conventional superconductivity (SC) and charge density wave. In a class of strongly-correlated materials, the interplay between electron correlations and electron-phonon interactions is believed to induce novel phenomena such as the unconventional high-$T_c$ $s$-wave SC in the alkali-doped fullerenes\cite{ganin2010,capone2002,nomura2015}. Even for high-$T_c$ cuprates, some experiments\cite{lanzara2001} and theoretical studies\cite{huang2003,ishihara2004,johnston2012} have suggested the important roles of phonons for full understanding the electronic properties including the SC. However, they are still controversial because the relevance of the electron-phonon interaction addressed in previous theoretical works largely rely on adjustable model parameters introduced in an {\it ad hoc} fashion. In addition, computationally accurate framework to study the interplay between the electron correlation and the electron-phonon interaction has not fully been explored. To establish the roles of phonons in a wide range of stongly correlated materials including the cuprates, we need a flexible method which can accurately treat strong electron-electron and electron-phonon interactions on an equal footing. 

For decades, variational Monte Carlo (VMC) methods have been applied to investigate strongly correlated electrons\cite{ceperley1997, yokoyama1990,capello2005}. Its advantage is that it does not suffer from the notorious negative-sign problem, whereas its accuracy depends on the assumed variational wave function. However, owing to the improved efficient optimization method such as the stochastic reconfiguration method\cite{sorella2001}, its accuracy and flexibility have improved by introducing many variational parameters\cite{sorella2005,tahara2008,kaneko2014,misawa2014,morita2015,kurita2015,tocchio2008,tocchio2011}. It has been recently applied to complicated {\it ab initio} multi-orbital effective Hamiltonians\cite{shinaoka2012, misawa2014_2,hirayama2015}. Recently, we have successfully extended this many-variable VMC (mVMC) method to electron-phonon coupled systems\cite{ohgoe2014}.

The Holstein-Hubbard model is the simplest model for studying the interpley of electron-electron and electron-phonon interactions. However, the phase diagram and physical properties under these two competing interactions are controversial even for the ground states. In one dimension and the Bethe lattice with inifinite coordination, its phase diagrams have been obtained by the density matrix renormalization group (DMRG)\cite{clay2005,tezuka2007,fehske2008} and the dynamical mean-field theory (DMFT)\cite{bauer2010,murakami2013}, respectively. At half filling, the DMRG studies have reported the existence of an intermediate metallic phase between a Mott insulating and a CO phase in the ground-state phase diagram. On the other hand, the DMFT study for zero temperature has not found its evidence\cite{bauer2010}. For square lattices, a finite-temperature quautum Monte Carlo (QMC) study has also suggested the emergence of an intermediate paramagnetic metallic phase between the AF and CO phases\cite{nowadnick2012,johnston2013}. However, such a phase diagram cannot be conclusive in the finite-temperature studies because of the Mermin-Wagner theorem.

Another important open issue is found when carriers are doped into the half-filled system. The DMFT study on the Holstein model has revealed the presence of a coexisting phase of CO and SC which is not prevented by the PS\cite{murakami2014}. It is interesting to ask whether the coexistence also exists in two dimensions. The connection between the SC and PS is also intriguing and has been discussed in the literature\cite{grilli1991,raimondi1993} for a different context of the three-band Hubbard model as a model for the cuprates. Recently, their strong connections are observed in the mVMC studies on the Hubbard model\cite{misawa2014} and {\it ab initio} effective Hamiltonian of electron-doped LaFeAsO\cite{misawa2014_2}. A natural question here is whether a phonon-driven PS also has a connection in the case of the $s$-wave SC. In this paper, we study these issues by using the mVMC method.

{\it Model}. ---The Hamiltonian we consider here is given by
\begin{eqnarray}
  {\cal H} & = & - t \sum_{\langle i,j \rangle, \sigma} (c_{i \sigma}^{\dagger}c_{j \sigma} + {\rm h. c.}) + U \sum_i n_{i \uparrow} n_{i \downarrow} \nonumber\\ 
  & & + g \sum_{i} x_{i} n_{i}  + \sum_i \left( \frac{p_i^2}{2M}  + \frac{M \Omega^2 x_i^2}{2} \right),
\end{eqnarray}
where $t$, $U$, $g$, and $\Omega$ represent the hopping amplitude, the on-site intraction strength between electrons, the electron-phonon interaction strength, and the phonon frequency, respectively. $c_{i \sigma}(c_{i \sigma}^{\dagger})$ represents the annihilation (creation) operator of an electron with spin $\sigma$ (=$\uparrow$ or $\downarrow$) at the site $i$. The particle number operators $n_{i \sigma}$ and $n_{i}$ are defined by $n_{i \sigma}=c_{i \sigma}^{\dagger} c_{i \sigma}$ and $n_i=n_{i \uparrow}+n_{i \downarrow}$. $x_i$ and $p_i$ are the lattice displacement operator and its conjugate meomentum operator, respectively. $x_i$ relates to the annihilation/creation boson(phonon) operator $b_{i}/b_{i}^{\dagger}$ as $x_i = \sqrt{\frac{1}{2M \Omega}} (b_{i} + b_{i}^{\dagger})$. The dimensionless electron-phonon interaction strength $\lambda$ is defined as the ratio of the lattice deformation energy to half the bandwidth $W/2=4t$ and we obtain $\lambda = g^2/(M \Omega^2 W)$, where $M$ is the mass of single-component nuclei. If we consider the path-integral representation of the partition function and integrate out the phonon degrees of freedom, the model is exactly mapped onto the Hubbard model with the effective dynamical on-site interaction $U_{\rm eff}(\omega)=U-\frac{\lambda W}{1-(\omega/\Omega)^2}$. In this paper, we set $M=t=1$ as the unit of mass and energy. We consider $N=L^2$ systems on the square lattice with $N_e$ electrons and impose the periodic/anti-periodic boundary condition in the $x$/$y$-direction to satisfy the closed-shell condition. The filling factor and doping (hole) concentration are given by $\rho=N_e/N$ and $\delta=1-\rho$, respectively.

{\it Method}. ---Our variational wave function takes the following form:$ | \psi \rangle = {\cal P}^{\rm el-ph} (| \psi^{\rm el} \rangle | \psi^{\rm ph} \rangle )$\cite{ohgoe2014}. Here, $| \psi^{\rm el} \rangle$ and $| \psi^{\rm ph} \rangle$ represent variational wave functions for electrons and phonons, respectively. ${\cal P}^{\rm el-ph}$ is the correlation factor which takes into account the entanglement between electrons and phonons. Its explicit form is given by $ {\cal P^{\rm el-ph}}  =  \exp \left( \sum_{i,j} {\alpha}_{ij} x_i n_j \right)$, where $\alpha_{ij}$ are variational parameters. 

As $| \psi^{\rm el} \rangle$, we adopt the generalized pairing wave function with the Gutzwiller\cite{gutzwiller1963} and Jastrow correlation factors\cite{jastrow1955}: $| \psi^{\rm el} \rangle = {\cal P}^{\rm J} {\cal P}^{\rm G} | \phi^{\rm pair} \rangle$. The generalized pairing wave function takes the form of $| \phi^{\rm pair} \rangle = \left( \sum_{i,j=1}^{N} f_{ij} c_{i \uparrow}^{\dagger} c_{j \downarrow}^{\dagger}  \right)^{N_{\rm e}/2} | 0 \rangle$, where $f_{ij}$ are variational parameters. This is a generalization of the Hartree-Fock-Bogliubov type wave funcion with AF/CO and SC orders\cite{tahara2008,giamarchi1991} and thus flexibly describes these states as well as paramagnetic metals (PM). In order to reduce the number of independent variational parameters, we assume that $f_{ij}$ have a sublattice structure such that $f_{ij}$ depend on the relative vector ${\bm r}_i-{\bm r}_j$ and a sublattice index of the site $i$ which we denote as $\eta(i)$. Thus, $f_{ij} = f_{\eta(i)}({\bm r}_i-{\bm r}_j)$. In the present study, we assume a 2 $\times$ 2 sublattice structure and the number of independent $f_{ij}$ reduces from $N^2$ to $2 \times 2 \times N$. We also assume a translational symmetry for variational parameters in the correlation factors. 

For $| \psi^{\rm ph} \rangle$, we use the tensor product of phonon wave functions with wave vectors ${\bm q}$: $| \psi^{\rm ph} \rangle  = \prod_{\bm q} | \psi^{\rm ph}_{\bm q} \rangle$. $| \psi^{\rm ph}_{\bm q} \rangle $ is expanded in terms of phonon Fock states $| m_{\bm q} \rangle$ as $| \psi^{\rm ph}_{\bm q} \rangle = \sum_{m_{\bm q}=0}^{m_{\bm q}^{\rm max}} c_{m_{\bm q}} | m_{\bm q} \rangle$. Here, $m_{\bm q}^{\rm max}$ are controllable cutoffs for the number of phonons and $c_{m_{\bm q}}$ are treated as variational parameters of real numbers. The number of its variational parameteres is $\sum_{\bm q} (m_{\bm q}^{\rm max}+1)$, which is equal to $N (m^{\rm max}+1)$ if we take $m_{\bm q}^{\rm max}=m^{\rm max}$. In this study, we checked the convergence of physical quantities as a function of the cutoff and we typically took $m_{\bm q}^{\rm max}=10-40$ for ${\bm q}=(\pi, \pi)$ and $m_{\bm q}^{\rm max}=5$ for others. As initial states in the optimization of variational parameters, we considered the non-interacting Fermi sea (PM state), SC, AF, CO, and coexisting states of SC+AF and SC+CO.

\begin{figure}[b]
\begin{center}
  \includegraphics[width=7cm]{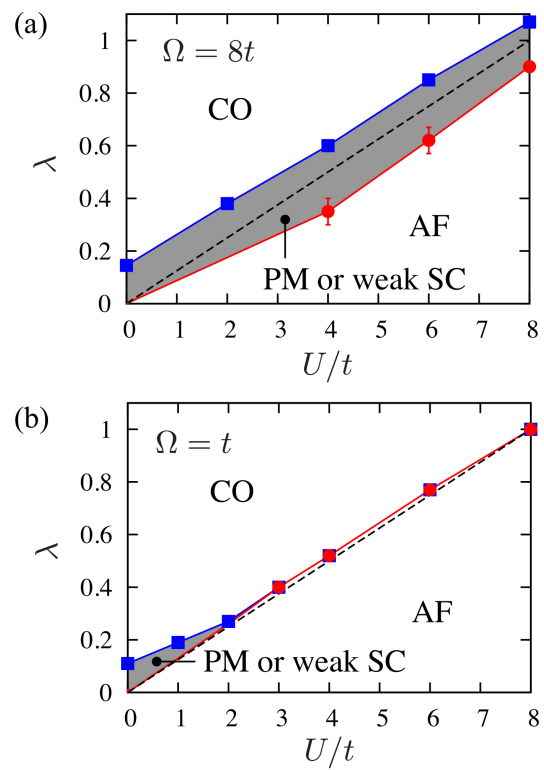}
\caption{(Color online) Ground-state phase diagram of the half-filled Holstein-Hubbard model on a square lattice at (a) $\Omega=8t$ and (b) $\Omega=t$. The blue squares and red circles represent the boundaries of CO and AF, respectively. Error bars are drawn, but most of them are smaller than the symbol size. Lines are used to guide the eye. Based on the fact that if $\lambda=0$, the system is an AF insulator for any $U>0$\cite{hirsch1985}, we put the starting point of the AF boundary at the origin. Shaded region represents the intermediate PM or weak SC phase.}\label{fig:phasediag_halffilled}\end{center}
\end{figure}

{\it Half-filled case}. --- We consider two phonon frequencies: An intermediate frequency $\Omega=8t$ (equal to the bandwidth $W$) and a smaller one $\Omega=t$. In Fig. {\ref{fig:phasediag_halffilled}}, we summarize our results in the ground-state phase diagram in the $U-\lambda$ plane. The phase diagram includes the boundary of the AF and CO phases. To distinguish each phase, we measured the spin structure factor $S_s({\bm q}) = \frac{1}{3 N} \sum_{i,j} \langle {\bm S}_i \cdot {\bm S}_j \rangle e^{i {\bm q}\cdot  ({\bm r}_i-{\bm r}_j)}$ and the charge structure factor $S_c({\bm q}) = \frac{1}{N} \sum_{i,j} (\langle n_i n_j \rangle - \rho^2)e^{i {\bm q} \cdot ({\bm r}_i-{\bm r}_j)}$. 

One of the main findings in this Letter is the existence of an intermediate phase sandwiched by the AF and CO phases around $U \sim \lambda W$. For $\Omega=8t$, we found a wide intermediate phase. For smaller frequency $\Omega=t$, it is narrowed but still exists for $U \lesssim 2$. The shrinkage of the intermediate region for small $\Omega$ was also observed in one\cite{clay2005,tezuka2007} and infinite\cite{murakami2013} dimensions. Previous QMC studies suggested the intermediate region at $U=5t$~\cite{nowadnick2012, johnston2013}. However, the wider intermediate region there is probably because their calculation is at finite temperature $T/t = 0.25$.

\begin{figure}[htbp]
\begin{center}
  \includegraphics[width=7cm]{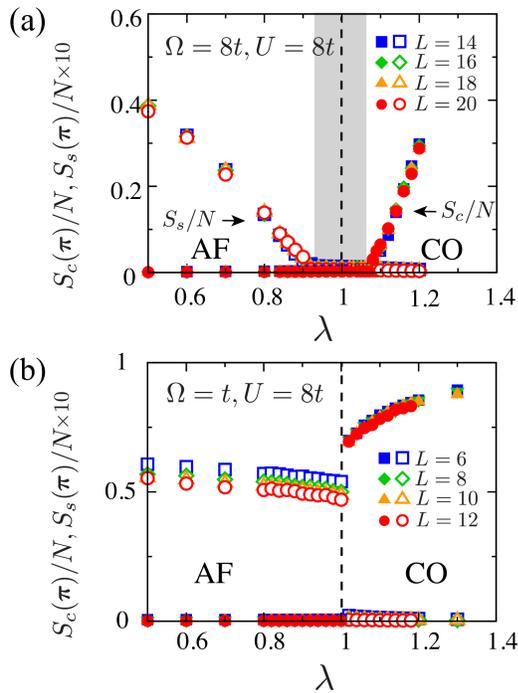}
\caption{(Color online) $S_{s}(\pi,\pi)/N$ and $S_{c}(\pi,\pi)/N$ as functions of $\lambda$ at (a) $(\Omega/t, U/t)=(8,8)$ and (b) (1,8), respectively. The vertical dashed line represents $U_{\rm eff}=0$. The shaded region indicates the intermediate metallic (weakly SC) phase. }\label{fig:phy_halffilled} 
\end{center}
\end{figure}

\begin{figure}[H]
\begin{center}
  \includegraphics[width=8.5cm]{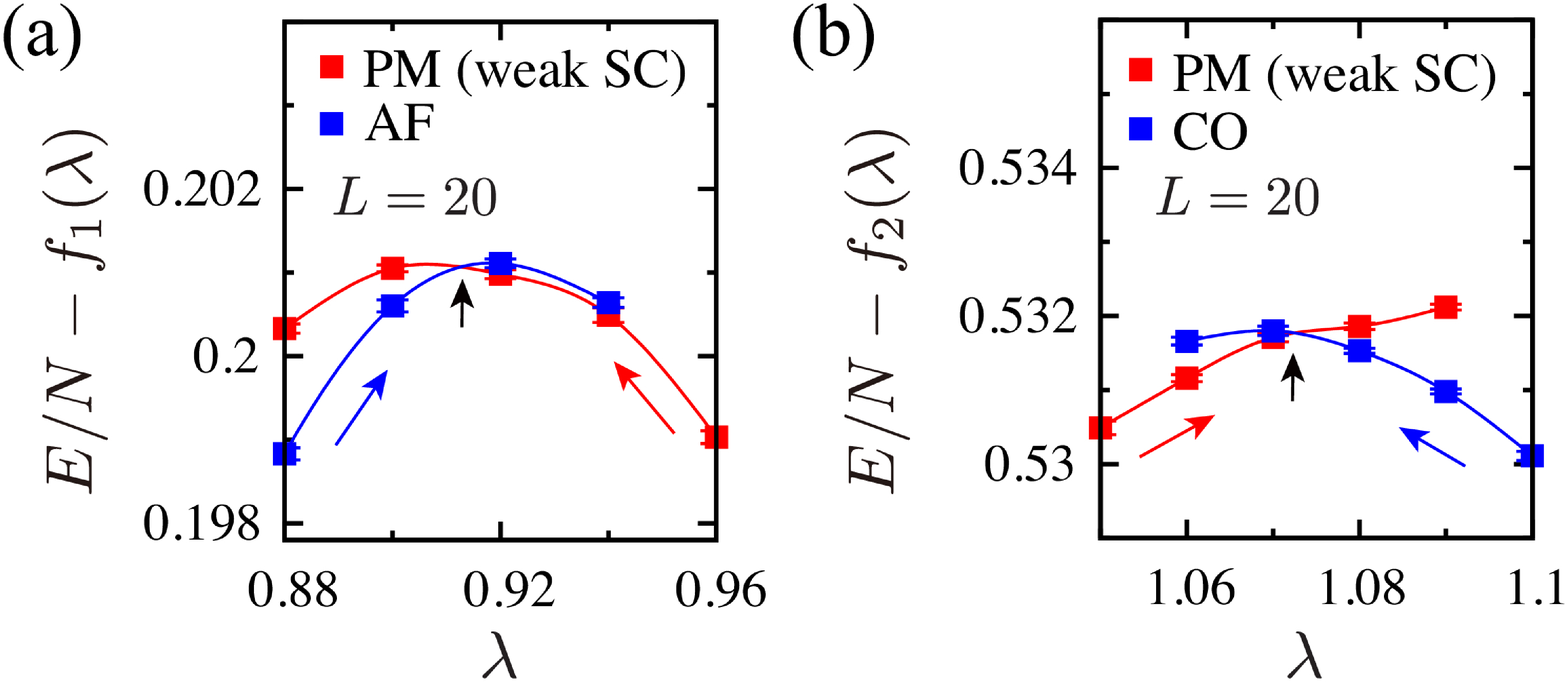}
\caption{(Color online) $E/N-\lambda$ curves of PM (weak SC) states crossing with (a)AF and (b)CO states at $(\rho, \Omega/t, U/t)=(1, 8, 8)$. The paths are along the vertical line at $U/t=8$ at the top right of Fig.~\ref{fig:phasediag_halffilled}(a). The curves are obtained by gradually changing $\lambda$ (in the direction indicated as colored arrows). Unimportant linear terms $f_1(\lambda)$ and $f_2(\lambda)$ $\propto \lambda$ are subtracted from $E/N$ for clarity. The crossing points are indicated as black arrows.}\label{fig:ene_crossing}\end{center}
\end{figure}

In Fig. {\ref{fig:phy_halffilled}} (a) and (b), we plot the spin/charge structure factor $S_{s/c}(\pi,\pi)/N$ as a function of $\lambda$ at $(\Omega/t,U/t)=(8,8)$ and (1,8), respectively. In the intermediate region, the values of $S_{s/c}(\pi,\pi)/N$ vanish after its extrapolation to the thermodynamic limit (see Supplemental Materials\cite{suppl} for the extrapolation procedure). The presence of the intermediate phase is further evidenced by the two first-order transitions signaled by two energy-level crossings as a function of $\lambda$, where the AF phase energy crosses with the intermediate phase energy at $\lambda \sim 0.91$ as shown in Fig.~\ref{fig:ene_crossing}(a) and then the latter crosses with the CO phase slightly above  $\lambda \sim 1.07$ as in Fig.~\ref{fig:ene_crossing}(b) with increasing $\lambda$ at fixed $U$ and $\Omega$. One may infer that the antiadiabatic or the adiabatic limits may further give useful insights. These are examined in Supplemental Materials\cite{suppl}. In the intermediate region, it is likely that weak SC orders emerge, while expected amplitudes of the order are too weak so that we could not distinguish them from PM states in the available data of finite systems (see Supplemental Materials\cite{suppl}).

\begin{figure}[htbp]
\begin{center}
  \includegraphics[width=8cm]{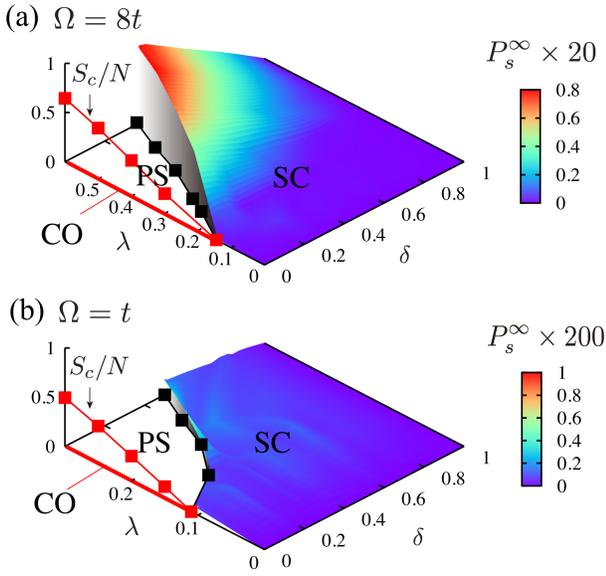}\caption{(Color online) Ground-state phase diagams of the Holstein model in the $\delta$-$\lambda$ plane at (a) $\Omega=8t$ and (b) $\Omega=t$. In the vertical axis, $S_{c}(\pi,\pi)/N$ (red squares) and $P_{s}^{\infty}$ (color plots) for $L=14$ are plotted in the CO and SC phase, respectively. Black squares in the bottom plane represent boundaries between PS and $s$-wave SC regions. White areas denote the PS regions. Thick red lines at $\delta=0$ indicate the CO phase.}\label{fig:phasediag_dope}\end{center}
\end{figure}

{\it Doped case}. --- We now study the doped region. In Fig. {\ref{fig:phasediag_dope}}, we first present our ground-state phase diagram at $U=0$ in the $\delta-\lambda$ plane for $\Omega=8t$ and $\Omega=t$, because the $U=0$ phase diagram captures an essential aspect. For $U=0$, the effective interaction $U_{\rm eff}(\omega)$ has negative parts for $\omega < \Omega$, which lead to $s$-wave SC states except for the gapped CO phase at half filling. In our phase diagram, the SC + CO phase is absent. Instead, the PS region appears adjacent to the CO phase at half filling. We find that for the smaller phonon frequency, the PS region is enlarged. In the Supplemental Material\cite{suppl}, we present the phase diagram in the adiabatic limit as the extreme case. In Fig. {\ref{fig:phasediag_dope}}, we also plot $S_{c}(\pi,\pi)/N$ and the long-range part of the $s$-wave SC correlation function $P_{s}^{\infty}$ which is defined by $P_{s}^{\infty} = \frac{1}{M} \sum_{\sqrt{2}L/4<|{\bm r}|} P_{s}({\bm r})$. Here, ${\bm r}$ is the relative position vectors belonging to $(-L/2, L/2]^2$ and $M$ is the number of vectors satisfying $\sqrt{2}L/4<|{\bm r}|<\sqrt{2}L/2$ and the SC function $P_{s}({\bm r})$ is defined by $ P_{s}({\bm r}) = \frac{1}{N} \sum_{{\bm r}_i} \langle \Delta_{s}^{\dagger} ({\bm r}_i) \Delta_{s} ({\bm r}_i+{\bm r}) \rangle$ with the order parameter $\Delta_{s}({\bm r}_i) = c_{{\bm r}_i \uparrow} c_{{\bm r}_i\downarrow}$.

\begin{figure}[htbp]
\begin{center}
  \includegraphics[width=8.5cm]{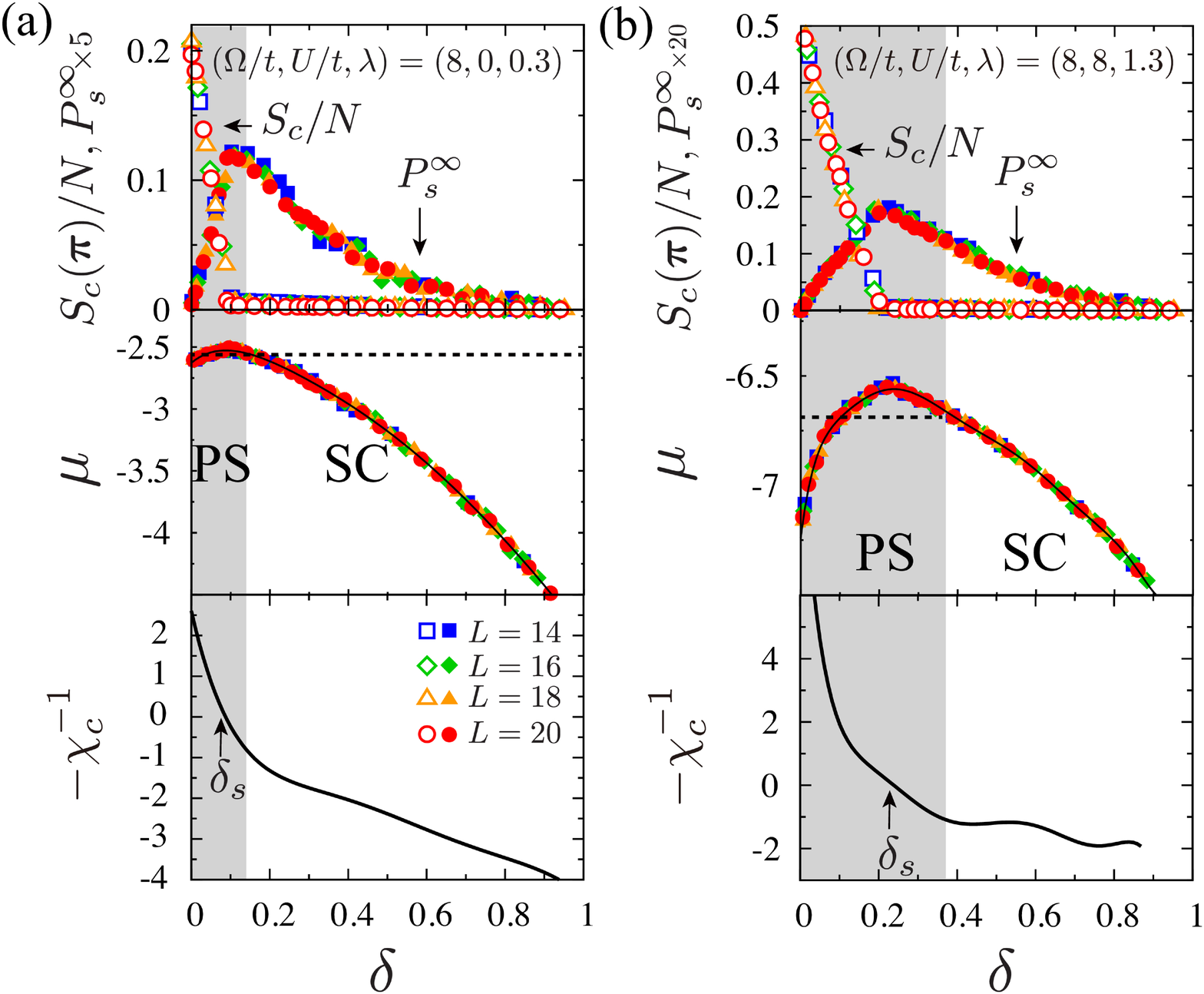}
\caption{(Color online) Physical quantities $S_{c}(\pi,\pi)/N$, $P_{s}^{\infty}$, $\mu$, and $-\chi_c^{-1}$ as functions of doping $\delta$ at (a) $(\Omega/t, U/t, \lambda) = (8, 0, 0.3)$ and (b) (8, 8, 1.3), respectively. The shaded area denotes the PS region which was determined by the Maxwell construction. The dashed horizontal line in the middle panel is used for the Maxwell construction. The curves of $-\chi_c^{-1}$ were derived from the derivative of the $\mu-\delta$ curves (black curves) which were obtained by the 7th order polynomial fit. The spinodal points $\delta_s$ are indicated as the arrows.}\label{fig:phy_dope}\end{center}
\end{figure}

In Fig.~{\ref{fig:phy_dope}}(a), we show physical quantities which were used to determine the phase diagrams in Fig.~\ref{fig:phasediag_dope} in an example at $(\Omega/t, U/t, \lambda) = (8, 0, 0.3)$.  We also show an interacting case for $(\Omega/t, U/t, \lambda) = (8, 8, 1.3)$ in Fig.~{\ref{fig:phy_dope}}(b) for comparison. Since in the antiadiabatic limit, the model is mapped to the standard Hubbard model with the on-site interaction $U_{\rm eff}=U-W\lambda$, the comparison between the interacting and noninteracting cases with the same $U_{\rm eff}$ may provide us with insight for large $\Omega$. The cases (a) and (b) indeed have the same $U_{\rm eff}=-2.4$. 
The value of $S_{c}(\pi,\pi)/N$ decreases monotonically and the CO eventually disappears at $\delta \simeq 0.14$ and 0.37 for (a)($U/t=0$) and (b)($U/t=8$), respectively. On the other hand, the value of $P_{s}^{\infty}$ increases as $\delta$ increases and we clearly observe SC phase. For small $\delta$, a CO and $s$-wave SC orders coexist. 
By the Maxwell construction for the $\delta$-$\mu$ curve, however, we find that the SC+CO phase is swallowed up by the PS region in our phase diagrams. Here, $\mu$ is the chemical potential which was calculated by $\mu({\bar N}_{\rm e}) = [ E(N_{\rm e}) - E(N'_{\rm e})]/(N_{\rm e}-N'_{\rm e})$. Here, $E$ is the total energy, ($N_{\rm e}$, $N'_{\rm e}$) are the electron numbers and we obtain the chemical potential at the mid filling ${\bar N}_{\rm e} = (N_{\rm e}+N'_{\rm e})/2$. Our Hamiltonian has the particle-hole symmetry at $\mu = -8\lambda -U/2$ = -2.4 and -6.4 for (a) and (b), respectively. Since this value is above the line used for the Maxwell construction, there is a charge gap at half filling. For the interacting case (b), the charge gap is even larger. We also present the negative inverse uniform charge susceptibility $-\chi_c^{-1} = \frac{d \mu}{d \rho}$ in Fig. {\ref{fig:phy_dope}}. 
In our model, the spinodal point $\delta_s$, where the uniform charge susceptibility diverges ($\chi_c^{-1}=0$), coincides with the critical point of the CO and therefore the PS is driven by the CO (see also results for the adiabatic limit in the Supplemental Material\cite{suppl}). 

Comparisons between (a) and (b) show a quantitative difference that the CO/SC orders are enhanced/supressed for large $U/t$. However, we find a universal common feature both in (a) and (b); a clear one-to-one correspondence among the peak of the SC order, the spinodal point and the border of the CO phase, thus indicates tight connections of the mechanism of the SC, CO and uniform charge instability. The strong effective attractive interaction of carriers is certainly a key, because it causes all of these three properties. The strong attraction is caused by the electron-phonon interaction here while the resultant charge fluctuations may also work as additional glue of the Cooper pair. The same trend between the enhancement of the $s$-wave SC and the uniform charge susceptibility has been reported for $d$-wave SC in the Hubbard model\cite{misawa2014} and extended $s$-wave SC in the {\it ab initio} effective Hamiltonian for LaFeAsO\cite{misawa2014_2} as well.

To summarize, by studying the ground states of the Holstein-Hubbard model on a square lattice, we have clarified where the $s$-wave SC is enhanced in the phase diagram. At half filling, we have found an intermediate metallic or weakly SC region sandwiched by the CO and AF phases. In the doped case, the SC is dramatically enhanced, but a wide PS region triggered by the CO largely hinders the SC and completely preempts the SC+CO phase. We have revealed that the SC is optimized at the border of the PS. These findings have been obtained by the VMC method extended for electron-phonon coupled systems. Our method is quite flexible, and therefore it will be also useful to study more complicated systems such as {\it ab initio} Hamiltonians of high $T_c$ cuprates where several different phonon modes are present.


We thank Kota Ido for useful discussions. T.O also thanks Yuta Murakami for discussions. The code was developed based on the open-source software mVMC\cite{mvmc}. This work is financially supported by the MEXT HPCI Strategic Programs for Innovative Research (SPIRE), the Computational Materials Science Initiative (CMSI) and Creation of New Functional Devices and High-Performance Materials to Support Next Generation Industries (CDMSI). This work was also supported by a Grant-in-Aid for Scientific Research (No. 22104010, No. 22340090 and No. 16H06345) from MEXT, Japan. The simulations were partially performed on the K computer provided by the RIKEN Advanced Institute for Computational Science under the HPCI System Research project (the project number hp130007, hp140215, hp150211, hp160201, and hp170263). The simulations were also performed on computers at the Supercomputer Center, Institute for Solid State Physics, University of Tokyo.

\bibliography{reference}

\newpage


{\LARGE Supplemental Material for \\ ``Competition among Superconducting, Antiferromagnetic, and Charge Orders with Intervention by Phase Separation in 2D Holstein-Hubbard Model''}
\beginsupplement

\section{Phase transitions to AF/CO phases at half-filling}
In Fig. {\ref{fig:nq_ext}, we show the extrapotaion of $S_{s/c}(\pi, \pi)/N$ to the thermodynamic limit. Here, we plot $S_{s/c}(\pi, \pi)/N$ as a function $1/L$ and performed a linear fit based on the spin-wave theory\cite{huse1988}. 

\begin{figure}[H]
\begin{center}
  \includegraphics[width=8.5cm]{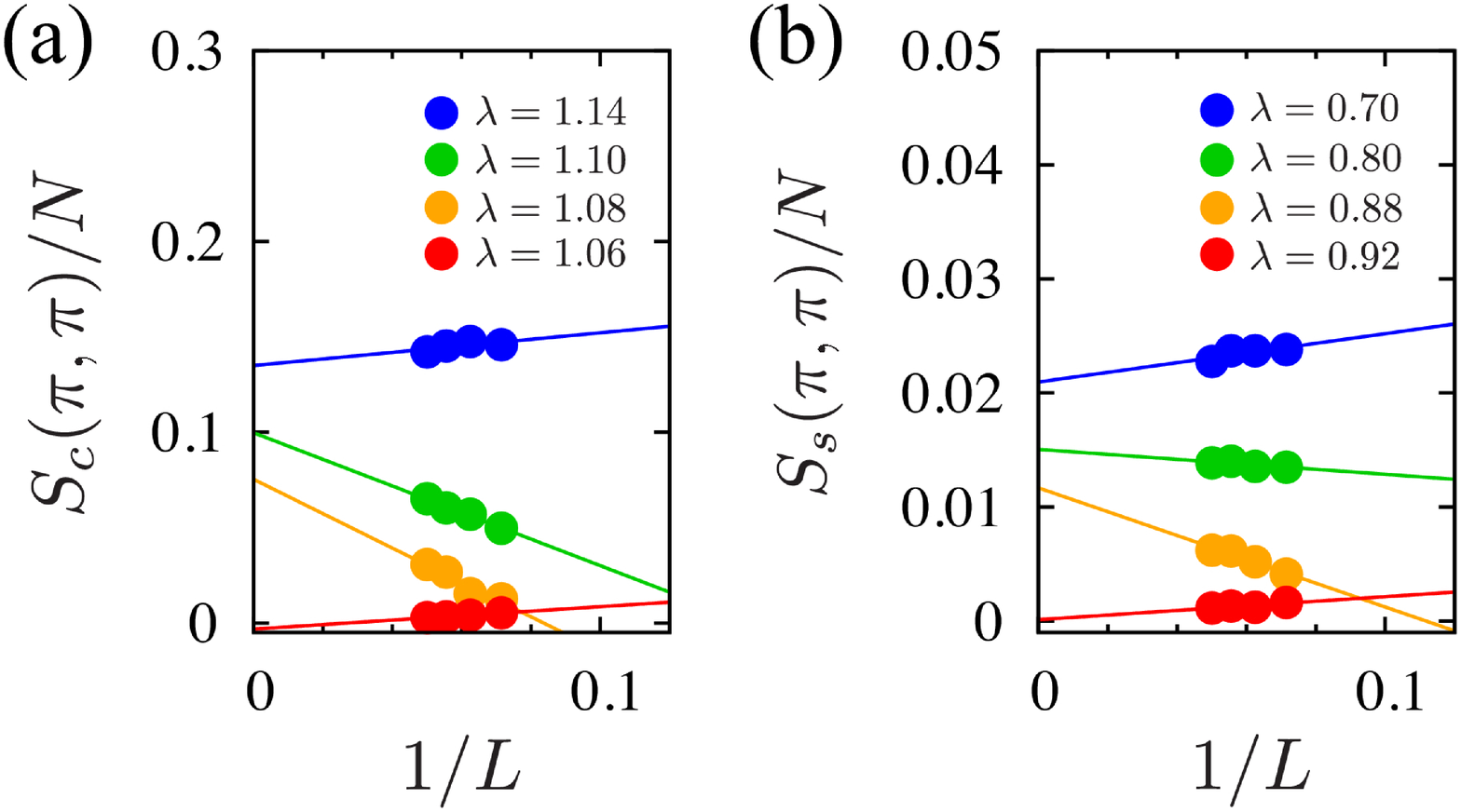}
\caption{(Color online) Extrapolations of (a) $S_s(\pi, \pi)/N$ and (b) $S_c(\pi, \pi)/N$ to the thermodynamic limit at $(\rho, \Omega/t, U/t)=(1, 8, 8)$.}\label{fig:nq_ext}\end{center}
\end{figure}

\section{Superconducting correlation function at half filling}
Here, we show the superconducting correlation functions at half filling and a small effective attraction $U_{\rm eff} (< 0)$ where the charge order does not appear. In Fig. {\ref{fig:sc_correl}}, we plot the $s$-wave superconducting correlation functions $P_{s}({\bm r})$ at $(U/t, \Omega/t, \lambda) = (0, 8, 0.14)$ $(U_{\rm eff}=-1.12t)$. $P_{s}({\bm r})$ does not show a clear power-law decay, but instead its long-range parts $P_{s}^{\infty}$ seem to show saturated behaviors. However, after the size extrapolation to the thermodynamic limit\cite{hurt2005} (the inset of Fig. {\ref{fig:sc_correl}}), its value does not exclude 0 within the error of $\sim 10^{-4}$. Because of this extrapolation error, we cannot estimate its value accurately, although it is expected to be finite.

Since our model is mapped onto the attractive Hubbard model with $U/t=-1.12$ in the antiadiabatic limit $\Omega/t \to \infty$ and it is equivalent to the repulsive Hubbard model with $U/t=1.12$ at half filling, we may estimate the upper bound of $P_s^{\infty}$ from the previous determinant quantum Monte Carlo (DQMC) results of the repulsive Hubbard model\cite{varney2009}. The antiferromagnetic order parameter defined by $m_{s} = \sqrt{\lim_{L \to \infty} S_s(\pi, \pi)/N}$ for the repulsive Hubbard model relates to $P_s^{\infty}$ for the attractive Hubbard model as $P_s^{\infty} = 2 m_{s}^2$ due to the spin-rotational symmetry. Since there are no available DQMC data of $m_{s}$ for $U/t<2$, we estimate the value at $U/t=1.12$ from the rescaled Hartree-Fock result. Here, we rescaled the Hartree-Fock result such that it reproduces the DQMC result at $U/t=2$ (shown in Fig. {\ref{fig:rescaledHF}}), although it seems to overestimate $m_s$ for $U/t<2$. Note that by considering the uncertainty in the fitting by the rescaled Hartree-Fock result, this should be regarded as a rough estimate. Nevertheless, from this estimation, we obtain $P_s^{\infty} \sim 0.001$ for the attractive Hubbard model with $U/t=-1.12$, whose order of estimate looks robust. (In the absence of the spin-rotational symmetry which is true for finite $\Omega/t$, we need to multiply the additional factor 1.5.) Thus, we confirm that the value of $P_s^{\infty}$ for $\Omega/t=8$ is smaller than the estimated value for the attractive Hubbard model. We believe that it is mainly due to the retardation effect. In any case, although the difficulty in the estimate of small order parameter exists, the order of expected superconducting order from this simple analysis suggests $P_s(r=\infty)\sim O(10^{-3}-10^{-4})$, from which our numerical results are consistent with the existence of a weak superconducting order. More quantitative analyses are left for future studies.

\begin{figure}[H]
\begin{center}
  \includegraphics[width=7.5cm]{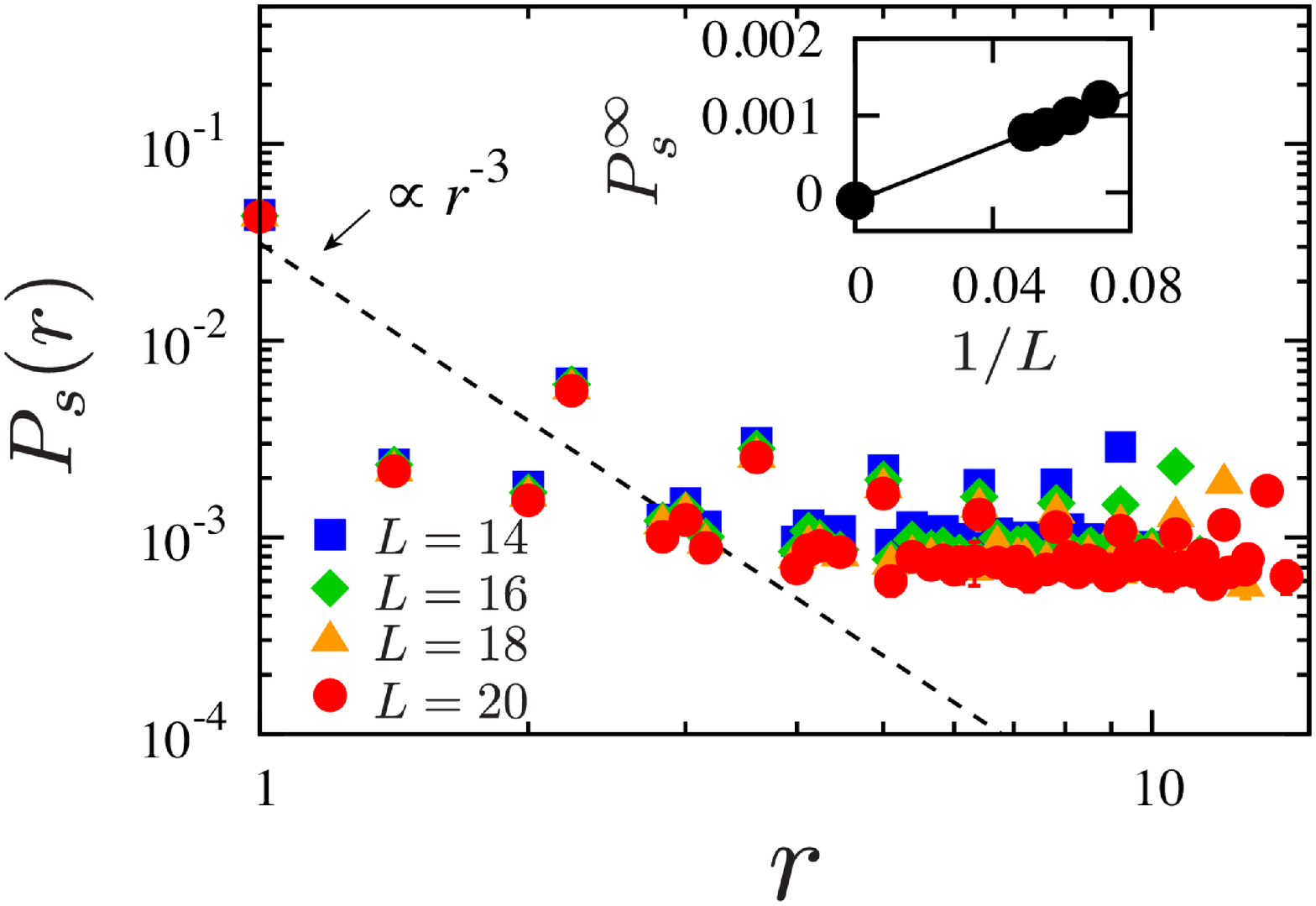}
\caption{(Color online) Logarithmic plots of $P_{s}({\bm r})$ at $(\Omega/t, U/t, \lambda)=(8, 0, 0.14)$. The dashed line represents the asymptotic $r^{-3}$ scaling for the non-interacting system\cite{furukawa1992}. In the inset, the extrapolation of $P_{s}^{\infty}$ to the thermodynamic limit is performed by a linear fitting.}\label{fig:sc_correl}\end{center}
\end{figure}

\begin{figure}[H]
\begin{center}
  \includegraphics[width=7cm]{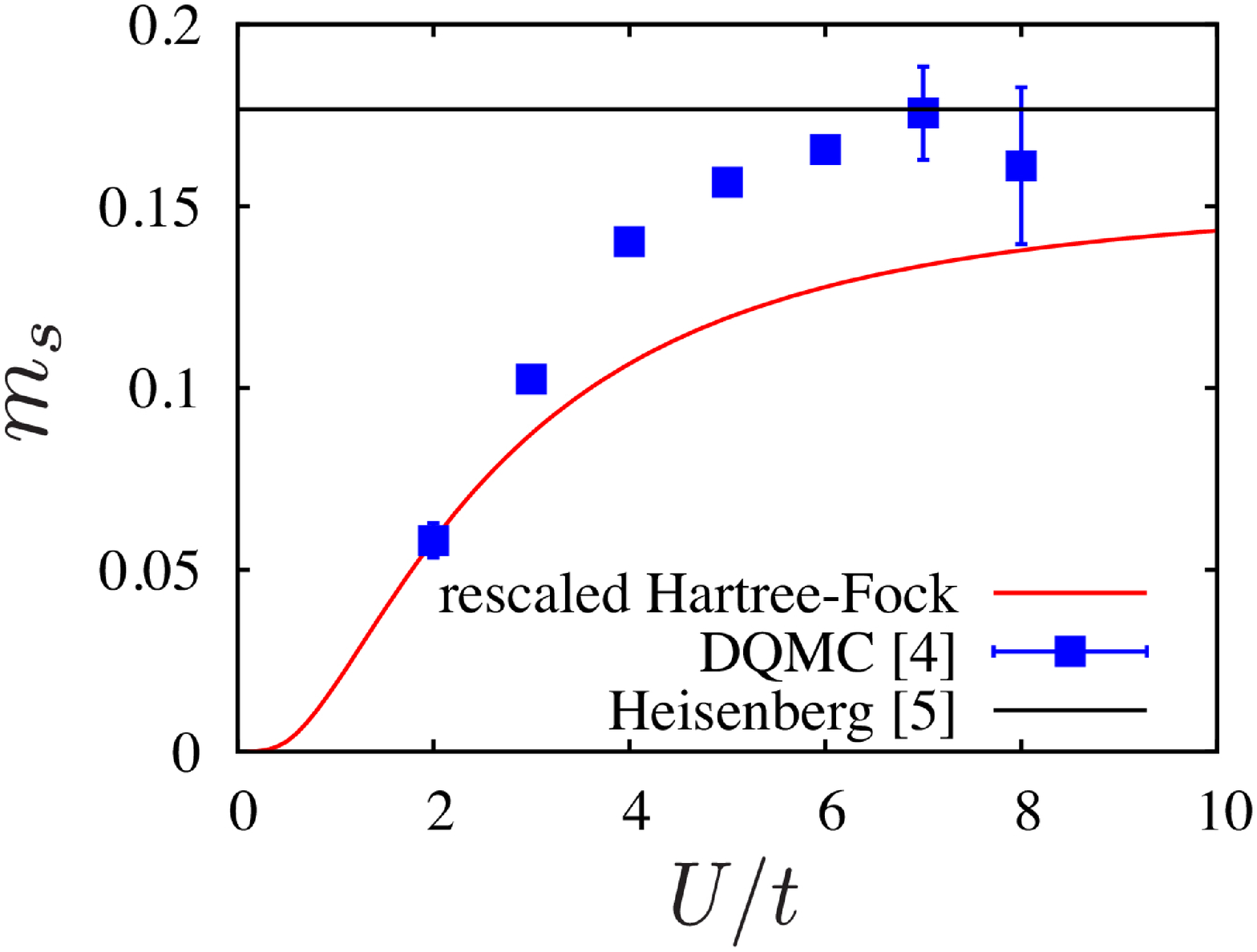}
\caption{(Color online) Rescaled Hartree-Fock results of $m_s$ (red curve). The DQMC results (blue squares) are taken from \cite{varney2009}. Note that our definition of the antiferromagnetic order parameter $m_s$ is smaller than theirs by factor $1/(2\sqrt{3})$ due to the difference in the definitions. The result for the Heisenberg model \cite{sandvik1997} is also shown as the horizontal black line.}\label{fig:rescaledHF}\end{center}
\end{figure}

\section{Antiadiabatic regime}
For $\Omega \gg t$, the Holstein-Hubbard model is reduced to the Hubbard model with the static interaction $U_{\rm eff}$. Here, we confirm this property by comparing results of these two models. 

Through the celebrated canonical transformation\cite{dichtel1971,shiba1972}, the attractive Hubbard model (AHM) at half filling is transformed into the repulsive Hubbard model (RHM) at the same filling. Owing to this transformation, the physical quantities of the AHM and the RHM can be mapped to each other. In particular, $S_c({\bm \pi})/4$ and the structure factor for the $s$-wave superconductivity $P({\bm 0})$ for the AHM are transformed to the spin structure factor of the $z$-direction component $S_s^z({\bm \pi})$ and the $x$-$y$ plane component $S_s^x({\bm \pi}) + S_s^y({\bm \pi})$ for the RHM, respectively. Here, $P({\bm q}) = \frac{1}{N} \sum_{i,j} \langle c_{i \uparrow}^{\dagger} c_{i \downarrow}^{\dagger}  c_{j \downarrow} c_{j \uparrow} \rangle e^{i {\bm q}\cdot ({\bm r}_i-{\bm r}_j)}$ and $S_s^{\alpha}({\bm q}) = \frac{1}{N} \sum_{i,j} \langle S_i^{\alpha} S_j^{\alpha} \rangle e^{i {\bm q}\cdot  ({\bm r}_i-{\bm r}_j)}$ ($\alpha$= $x,y,z$).

The results are shown in Fig. {\ref{fig:suppl_stfactors}}. We plot $S_s({\bm \pi})/N$ and@$[S_c({\bm \pi}) + 4P({\bm 0})]/12N$ at $(\rho, U/t) = (1, 8)$ for large $\Omega$s. The $\Omega \to \infty$ results are obtained by the calculations for the Hubbard model with the effective interaction $U_{\rm eff}$. In this limit, these two quantities should be symmetric with respect to $U_{\rm eff}=0$ if the results are exact. For the Holstein-Hubbard model, we show results for $\Omega/t$=50, 100, and 200 at $U/t=8$. From these results, we can confirm that the results of the Holstein-Hubbard model approach those of the Hubbard model with $U_{\rm eff}$ for $\Omega \gg t$. In this limit, there is no intermediate region between the AF and CO phase. To show the absence of energy crossings in this limit, we present the energy curves for the Hubbard model in Fig. \ref{fig:enecurve_hm}.

\begin{figure}[H]
\begin{center}
  \includegraphics[width=7cm]{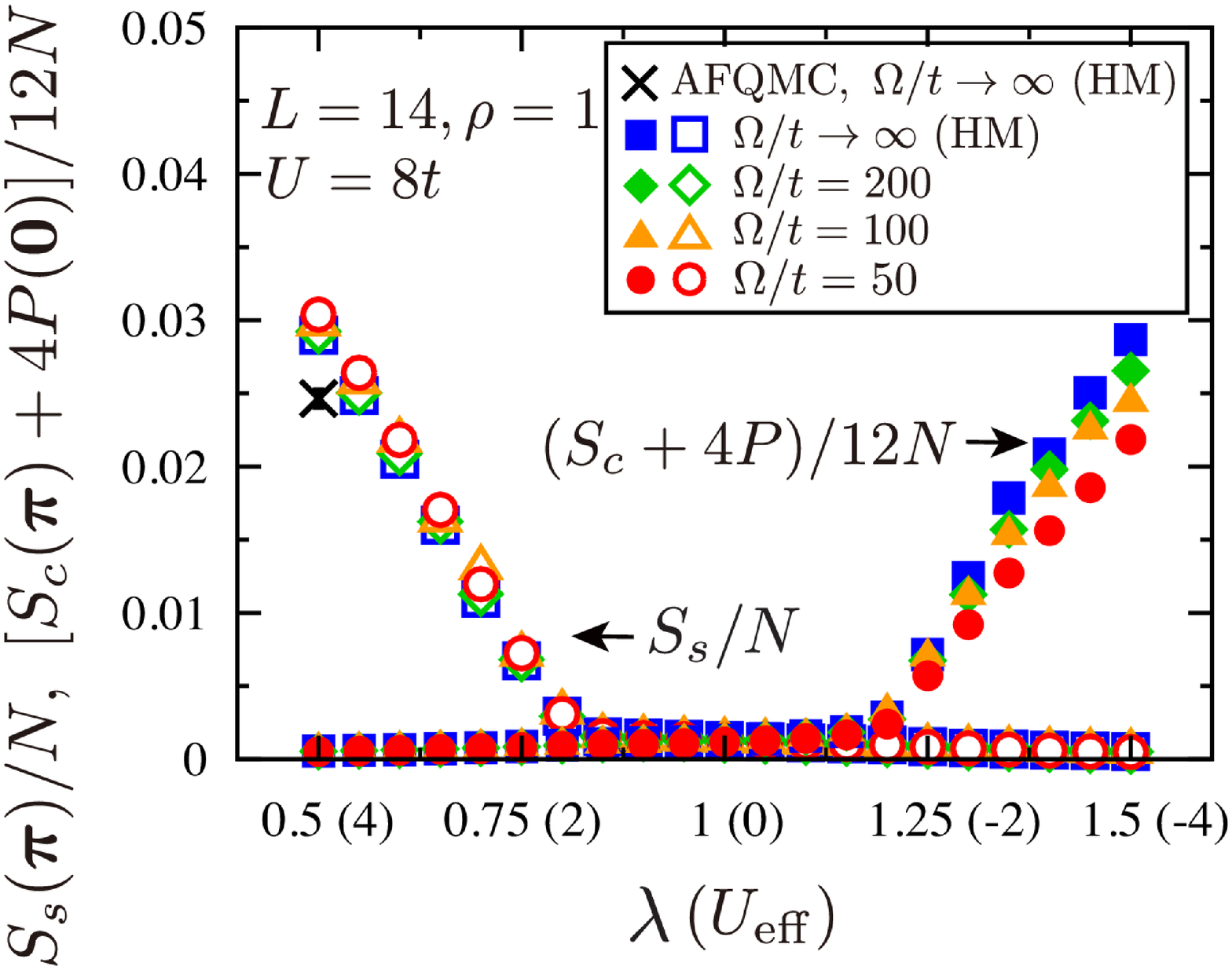}
\caption{(Color online) Structure factors $S_s({\bm \pi})/N$ (open symbols) and $[S_c({\bm \pi}) + 4P({\bm 0})]/12N$ filled symbols) for large $\Omega$s. The results for infinite $\Omega$ are obtained by calculations of the Hubbard model (HM) with the interaction $U_{\rm eff}$. The parenthesis in the horizontal axis represent the value of $U_{\rm eff}$. The linear system size is $L=14$. The value of $S_s({\bm \pi})/N$ from the auxiliary-field quantum Monte Carlo (AFQMC) method \cite{tahara2008} is shown for reference.}\label{fig:suppl_stfactors}\end{center}
\end{figure}

\begin{figure}[H]
\begin{center}
  \includegraphics[width=6.5cm]{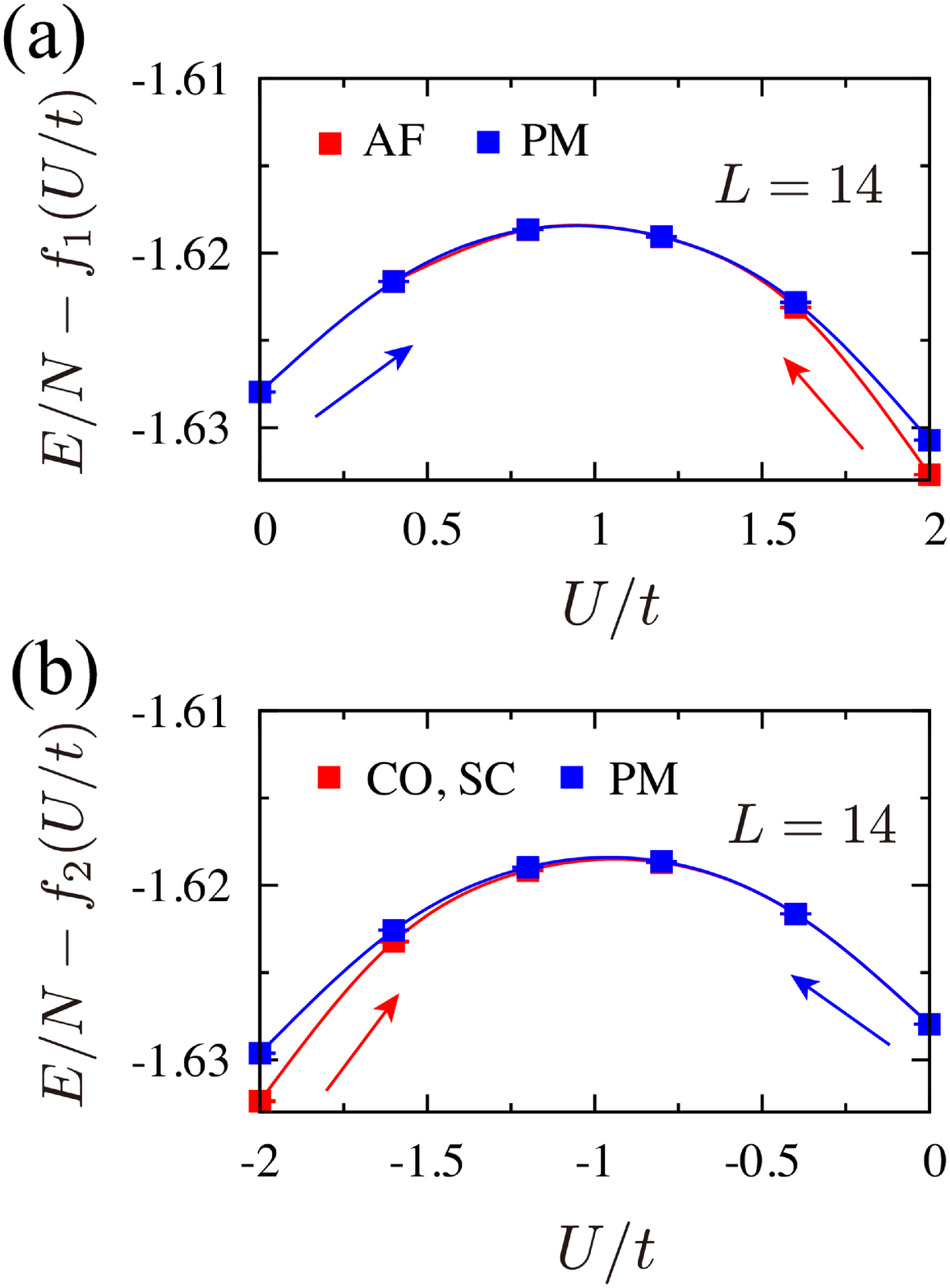}
\caption{(Color online) $E/N - U/t$ curves for the half-filled Hubbard model with the interaction $U$. The blue energy curves (denoted as PM) are obtained by gradually increasing the interaction $|U|$ from the PM state at $U=0$. The red energy curves (denoted as AF/CO,SC) are obtained by decreasing $|U|$ from the ground state at $U=2$/-2. Linear terms $f_1(U/t)$ and $f_2(U/t) \propto U/t$ are subtracted from $E/t$ for clarity.
We do not find any energy crossing within the numerical accuracy.
}\label{fig:enecurve_hm}\end{center}
\end{figure}

Because of the $SU(2)$ spin-rotational symmetry, there is the degeneracy between a SC and CO state for the half-filled attractive Hubbard model. To see how accurately we can describe this degeneracy, we here compare these two states. In Fig. {\ref{fig:degeneracy}}, we present the charge/superconducting structure factors ($L=14$). The charge/superconducting structure factors are enhanced for large $-U/t$ in the SC/CO states. We obtained these states by optimizing the initial states with SC or CO orders. We define the energy difference between these states as $\Delta E = (E_{\rm CO} -E_{\rm SC})/N$, where $E_{\rm CO}$ and $E_{\rm SC}$ are the energies of SC and CO states, respectively. We obtained $\Delta E/t = 0.00264(6)$ for $U/t=-4$. Within this difference ascribed to the error in the VMC calculations, the two states are nearly degenerate. The difference is quite small compared with the Holstein-Hubbard model where there is no degeneracy. For instance, we obtained $\Delta E/t = -0.1019(2)$ at $\Omega/t=U/t=8$ and $\lambda=1.5$ $(U_{\rm eff}/t=-4)$.

\begin{figure}[H]
\begin{center}
  \includegraphics[width=7.5cm]{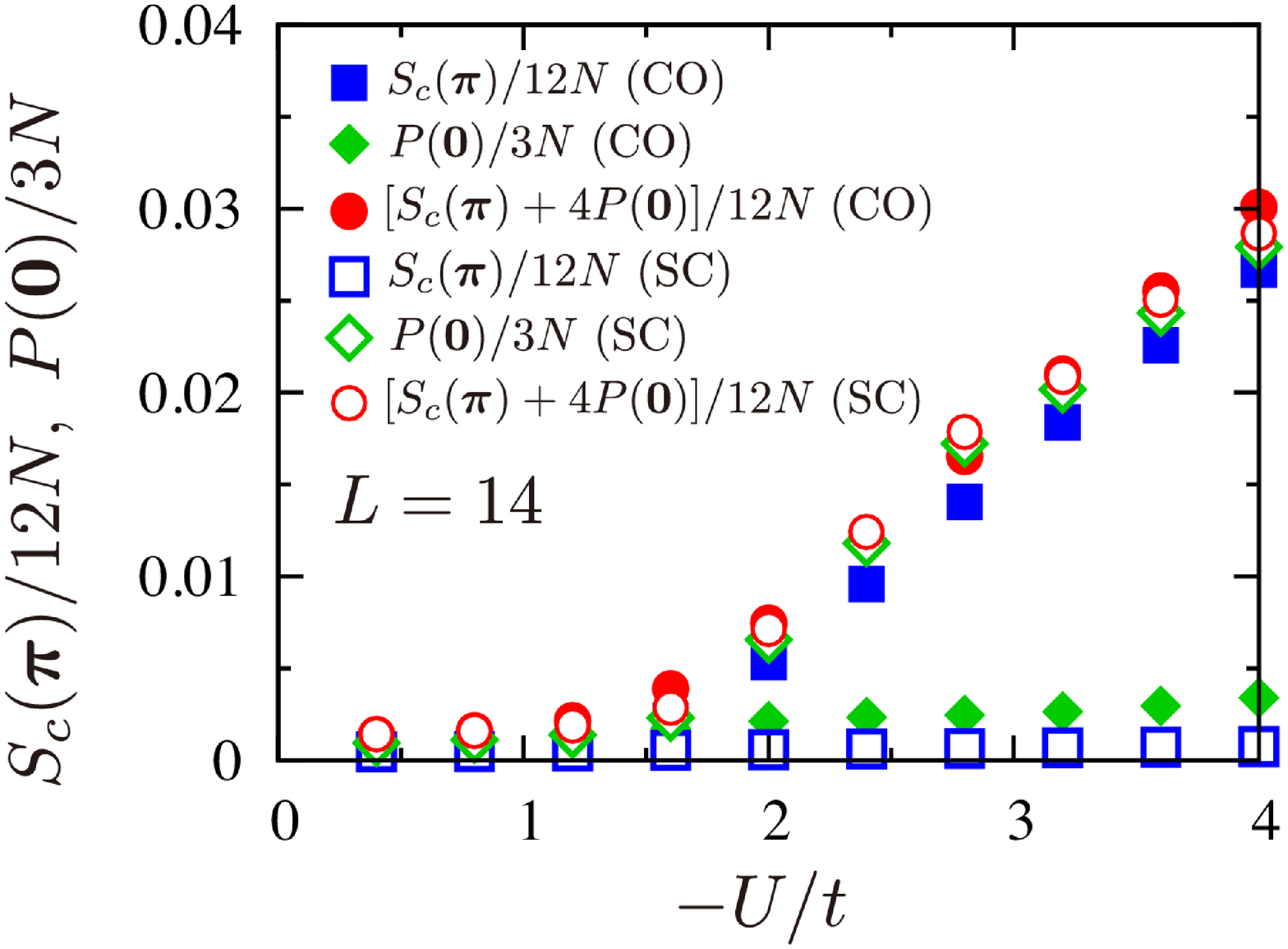}
\caption{(Color online) Comparison between SC and CO states for the half-filled attractive Hubbard model at $L=14$. In addition to $S_c({\bm \pi})/12N$ and $P({\bm 0})/3N$, we plot their sums which are mapped to $S_s({\bm \pi})/N$ through the canonical transformation.}\label{fig:degeneracy}\end{center}
\end{figure}

Finally, we present results for the doped case. In the main text, we showed the presence of phase separated regions for finite $\Omega/t$. In contrast, a QMC study reported the absence of phase separations for the static attractive Hubbard model\cite{moreo1991}. In addition, the superconductivity becomes dominant away from half filling and there is no degeneracy or coexistence with the charge order. To show that our VMC method can describe these properties in the antiadiabatic limit, we present physical quantities as functions of $\delta$ for the Hubbard model at $U/t=-2.4$ as well as the Holstein model at $\lambda=0.3$ ($U_{\rm eff}/t = -2.4$) and $\Omega/t=200$ in Fig. {\ref{fig:ahm_dope}}. From the good agreement between these two cases, we can again confirm that the Holstein model approaches its effective static Hubbard model for large $\Omega/t$. These results can be regarded as the antiadiabatic limit of Fig. 5 (a) where a phase separation was observed for $\Omega/t = 8$. In contrast, the monotonic behavior of $\mu$ here indicates the absence of phase separations in the antiadiabatic limit. We can also see that superconducting states are indeed dominant away from half filling in comparison to the degeneracy with the charge ordered state shown in Fig. {\ref{fig:degeneracy}}.  More concretely, we obtained $\Delta E/t = 0.0004(4)$ for $U/t=-2.4$ at half filling, whereas we found that the SC states always have lower energies than the CO states away from half filling.

\begin{figure}[H]
\begin{center}
  \includegraphics[width=7.5cm]{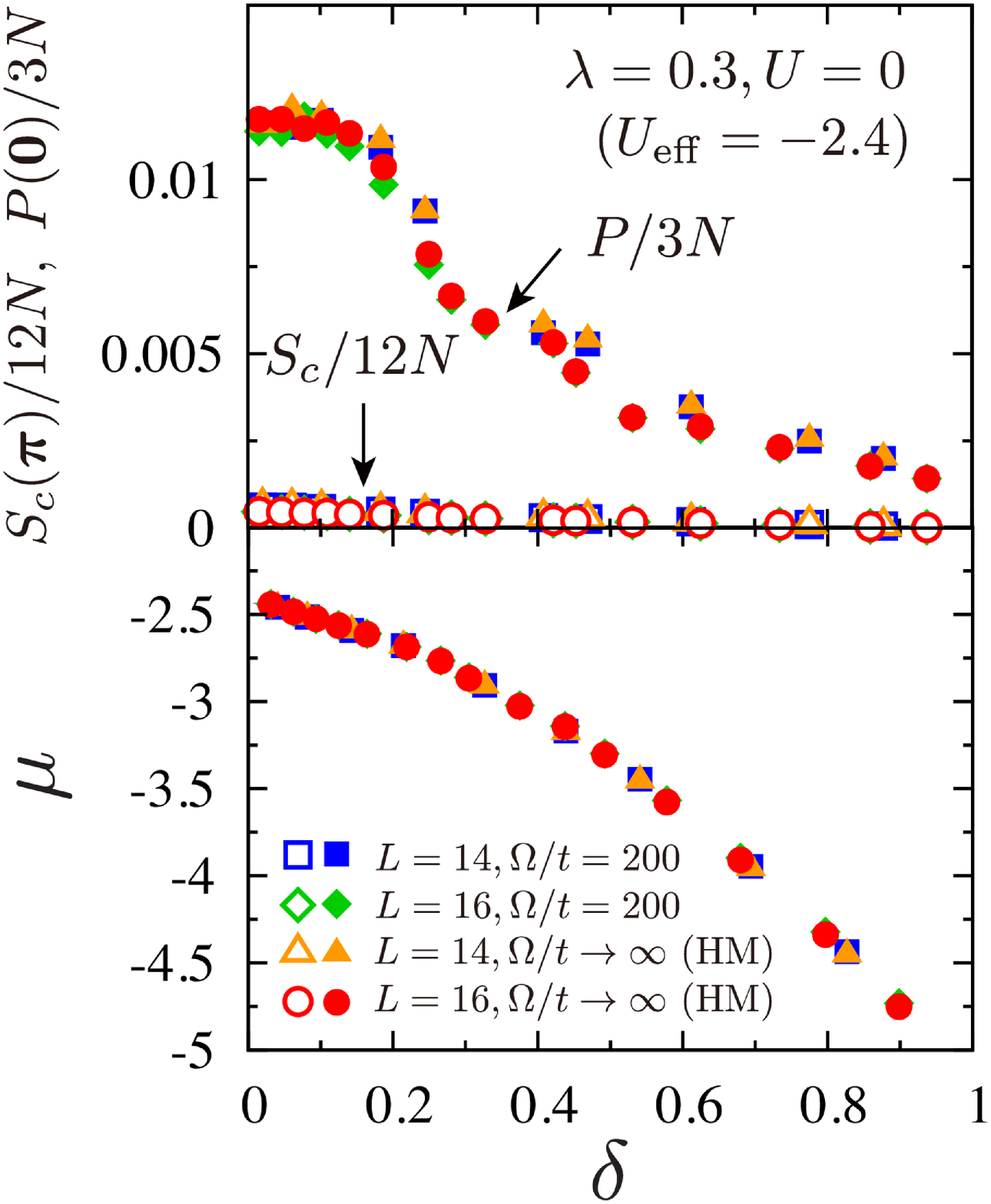}
\caption{(Color online) Physical quantities $S_c({\bm \pi})/12N$, $P({\bm 0})/3N$ and $\mu$ as functions of doping $\delta$ for the Holstein model at $\lambda = 0.3$ for $\Omega/t=200$ and its antiadiabatic limit $\Omega/t \to \infty$. In the latter case, we treated the Hubbard model (HM) with the effective interaction $U_{\rm eff} = -2.4$. For comparison, an unimportant constant 1.2 is subtracted for $\mu$ of the HM. }\label{fig:ahm_dope}\end{center}
\end{figure}

\section{Doped Holstein model in the adiabatic regime}
In Fig. 4, we have shown the phase diagrams for the doped Holstein model at $\Omega/t=1$ and 8. The physical quantities for $\Omega/t=1$ and $\lambda=0.3$ are presented in Fig. {\ref{fig:suppl_dope_omega1}}. From these results, we found that for smaller phonon frequencies, the PS region is enlarged. As the extreme case, we here present the analysis for the adiabatic limit.

\begin{figure}[b]
\begin{center}
  \includegraphics[width=6cm]{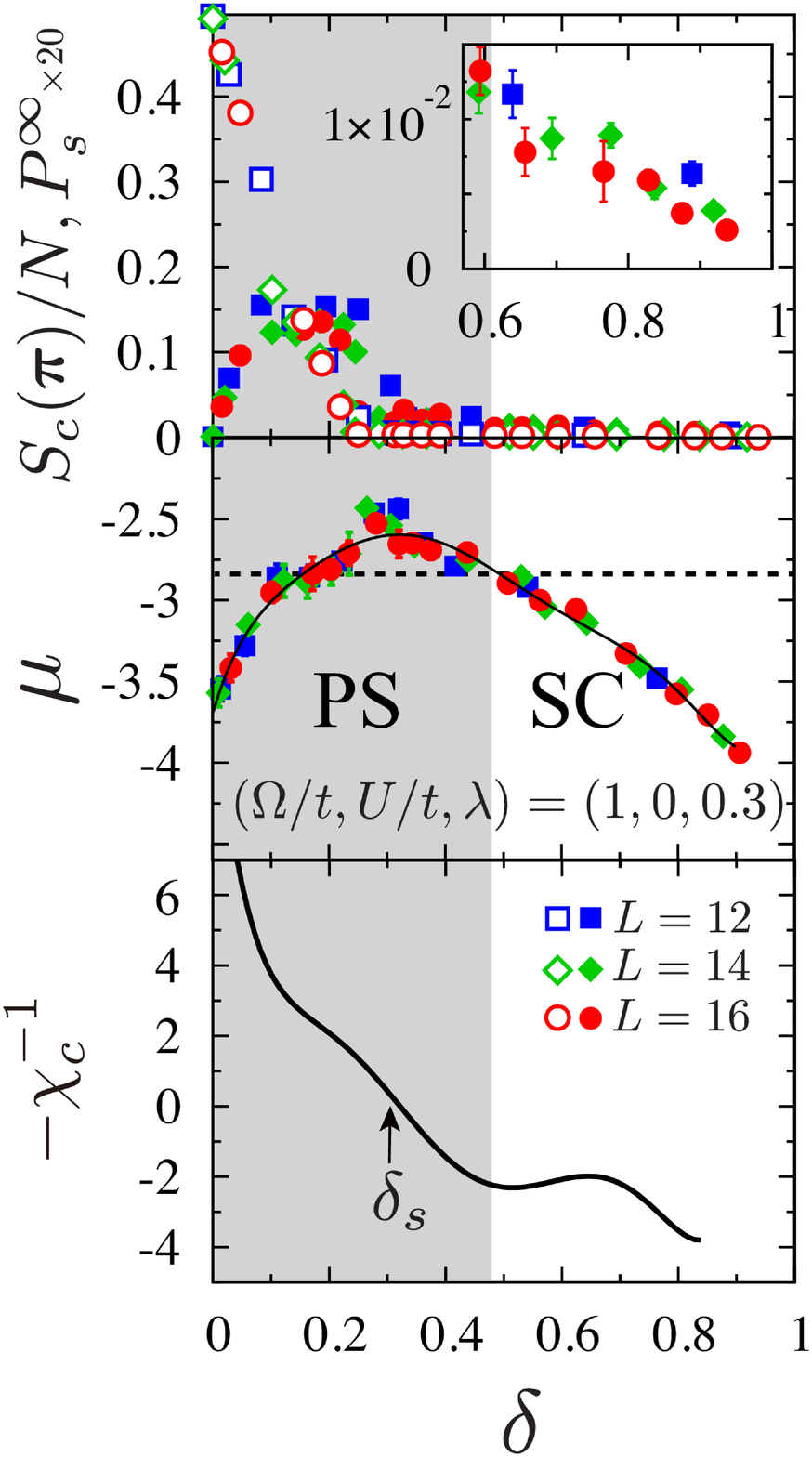}
\caption{(Color online) Physical quantities $S_{c}(\pi,\pi)/N$, $P_{s}^{\infty}$, $\mu$, and $-\chi_c^{-1}$ as functions of doping $\delta$ at $(\Omega/t, U/t, \lambda) = (1, 0, 0.3)$. }\label{fig:suppl_dope_omega1}\end{center}
\end{figure}

The limit of $\Omega/t \to 0$ with the spring constant $K = M\Omega^2$ kept fixed ($M \to \infty$) is called adiabatic (classical) limit. In this limit, the Hamiltonian is reduced to 
\begin{eqnarray}
  {\cal H} & = & - t \sum_{\langle i,j \rangle, \sigma} (c_{i \sigma}^{\dagger}c_{j \sigma} + {\rm h. c.}) + g \sum_{i} x_{i} n_{i}  \nonumber \\
  & & + \sum_i \frac{K x_i^2}{2}, 
\end{eqnarray}
where $K$ is the spring constant and the lattice displacements $\{x_i\}$ become classical variables. By changing variables as $x_i = \tilde{x}_i -\frac{g}{K} \rho$, we can rewrite it as
\begin{eqnarray}
  {\cal H} & = & - t \sum_{\langle i,j \rangle, \sigma} (c_{i \sigma}^{\dagger}c_{j \sigma} + {\rm h. c.}) + g \sum_{i} \tilde{x}_{i} \delta n_{i} \nonumber \\
  & & + \sum_i \frac{K \tilde{x}_i^2}{2} - \frac{W \lambda}{2} \rho^2 N. 
\end{eqnarray}
Here, $\delta n_{i} = n_i - \rho$ and $\lambda = \frac{g^2}{KW}$. In this form, the new lattice displacements $\{ \tilde{x}_i \}$ are 0 if electrons are uniformly distributed ($\langle \delta n_{i} \rangle=0$ for all $i$'s). 

By completing the square for the second and third terms, the total energy $\langle {\cal H} \rangle$ can be minimized if $\tilde{x}_i = - \frac{g}{K} \langle \delta n_{i} \rangle$. We assume a charge orderd state with $\tilde{x}_{i} = (-1)^i \tilde{x} = - \frac{g}{K} \delta n (-1)^i$, where $\tilde{x}$ and $\delta n$ are order parameters. By substituting this for the Hamiltonian, we obtain
\begin{eqnarray}
  {\cal H} & = & - t \sum_{\langle i,j \rangle, \sigma} (c_{i \sigma}^{\dagger}c_{j \sigma} + {\rm h. c.}) + g \sum_{i} \tilde{x} (-1)^i n_{i} \nonumber \\
  & &+ \frac{K \tilde{x}^2}{2}N - \frac{W \lambda}{2} \rho^2 N. 
\end{eqnarray}
We can diagonalize this by the following unitary transformation:
\begin{eqnarray}
a_{{\bm k} \sigma}^{\dagger} & = & u_{{\bm k}} c_{{\bm k} \sigma}^{\dagger} + v_{{\bm k}} c_{{\bm k}+{\bm Q} \sigma} \\
b_{{\bm k} \sigma}^{\dagger} & = & -v_{{\bm k}} c_{{\bm k} \sigma}^{\dagger} + u_{{\bm k}} c_{{\bm k}+{\bm Q} \sigma}
\end{eqnarray}
with
\begin{eqnarray}
u_{{\bm k}} (v_{{\bm k}}) & = & \frac{1}{2} \left[ 1 -(+) \frac{\varepsilon ({\bm k})}{\sqrt{\varepsilon ({\bm k})^2 + \Delta^2}} \right].
\end{eqnarray}
Here, ${\bm Q}=(\pi, \pi)$. By this transformation, we obtain
\begin{eqnarray}
  {\cal H} & = & \sum_{{\bm k} \in {\rm \ folded\ BZ}, \sigma} [E^{-}({\bm k}) a_{{\bm k} \sigma}^{\dagger} a_{{\bm k} \sigma} + E^{+}({\bm k}) b_{{\bm k} \sigma}^{\dagger}b_{{\bm k} \sigma}] \nonumber \\
 & & +  \frac{K \tilde{x}^2}{2}N - \frac{W \lambda}{2} \rho^2 N,
\end{eqnarray}
with 
\begin{eqnarray}
  E^{\pm}({\bm k}) = \pm \sqrt{\varepsilon({\bm k})^2 + \Delta^2} 
\end{eqnarray}
Here, BZ denotes the Brillouin zone, $\varepsilon ({\bm k})$ is the energy dispersion for the non-interacting system and $\Delta = g \tilde{x}$. Based on this result, we can numerically calculate the total energy as a function of $\tilde{x}$ (or $\delta n$) and obtain the ground-state energy as the minimum of $E(\tilde{x})$.

In Fig. {\ref{fig:phasediag_adiabatic}, we present the obtained ground-state phase diagram. In the phase diagrm for finite $\Omega$ (Fig. 4 of the main text), we observe that the phase separation region expands as $\Omega$ decreases. Especially for large $\lambda$, a broad phase separation region is reasonable, because we have the term $- \frac{W \lambda}{2} \rho^2 N$ in the Hamiltonian. The second derivative of $- \frac{W \lambda}{2} \rho^2$ with respect to $\rho$ gives the negative constant $- W \lambda$ and this easily makes the curve of $E/N$ convex upward for large $\lambda$. As is evident, the origin of the term $- \frac{W \lambda}{2} \rho^2 N$ is the uniform shift of the original lattice displacement due to the change of the particle density. 

As seen for finite $\Omega$, the spinodal point $\delta_s$ coincides with the critical point of the charge order in the adiabatic limit as well. Actually this is not accidental, because the vanishing $\Delta$ makes a kink in the chemical potential. To show this, we plot the $\delta$-dependence of $\Delta$ and $\mu$ in Fig. {\ref{fig:phy_adiabatic}. This fact indicates that the charge order in the doped system is necessarily preempted by the phase separation.

\begin{figure}[H]
\begin{center}
  \includegraphics[width=7cm]{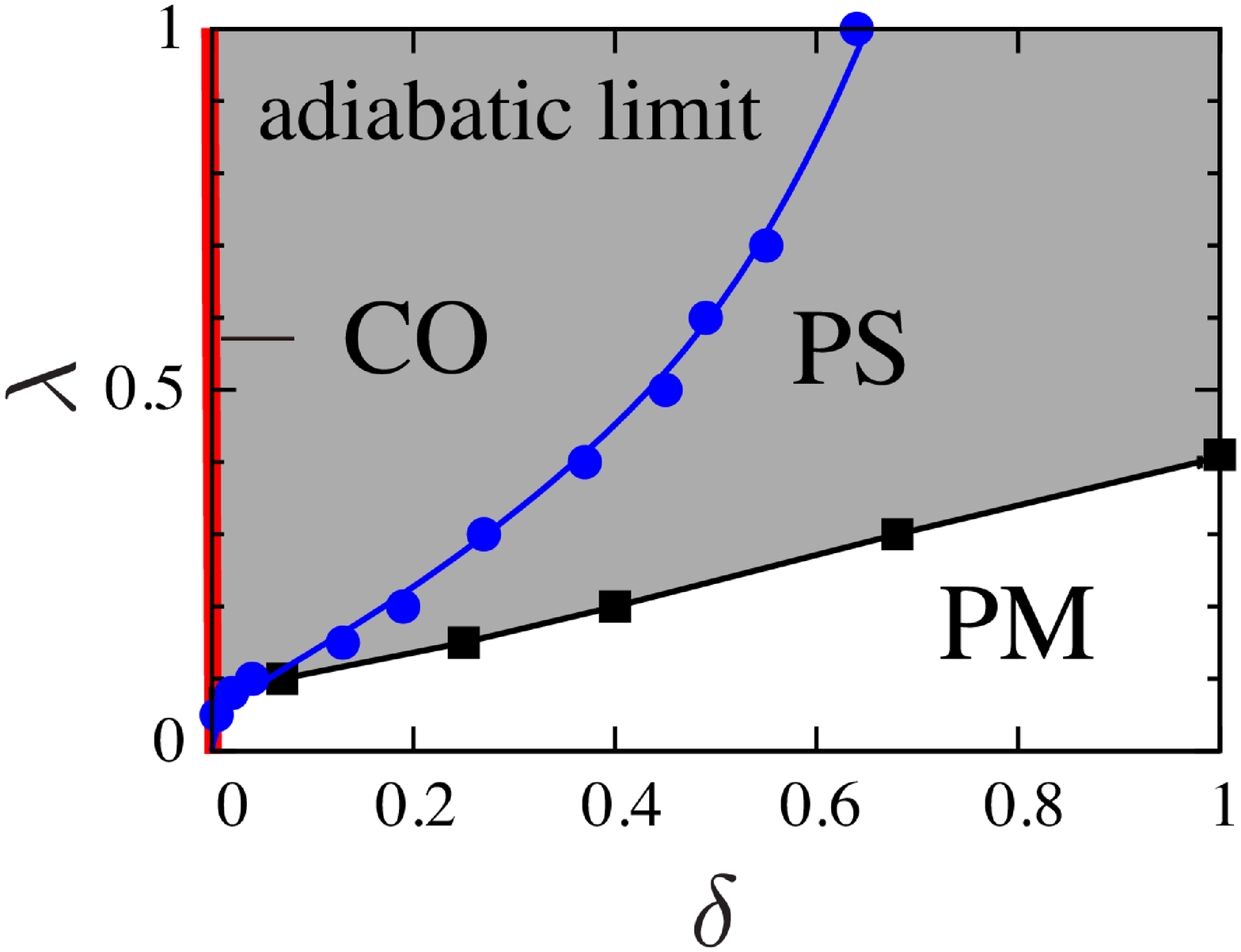}
\caption{(Color online) Ground-state phase diagram of the Holstein model (for a $40 \times 40$ system) in the adiabatic limit. Circles and squares represents the border of the charge order and phase separation, respectively. }\label{fig:phasediag_adiabatic}\end{center}
\end{figure}

\begin{figure}[H]
\begin{center}
  \includegraphics[width=9cm]{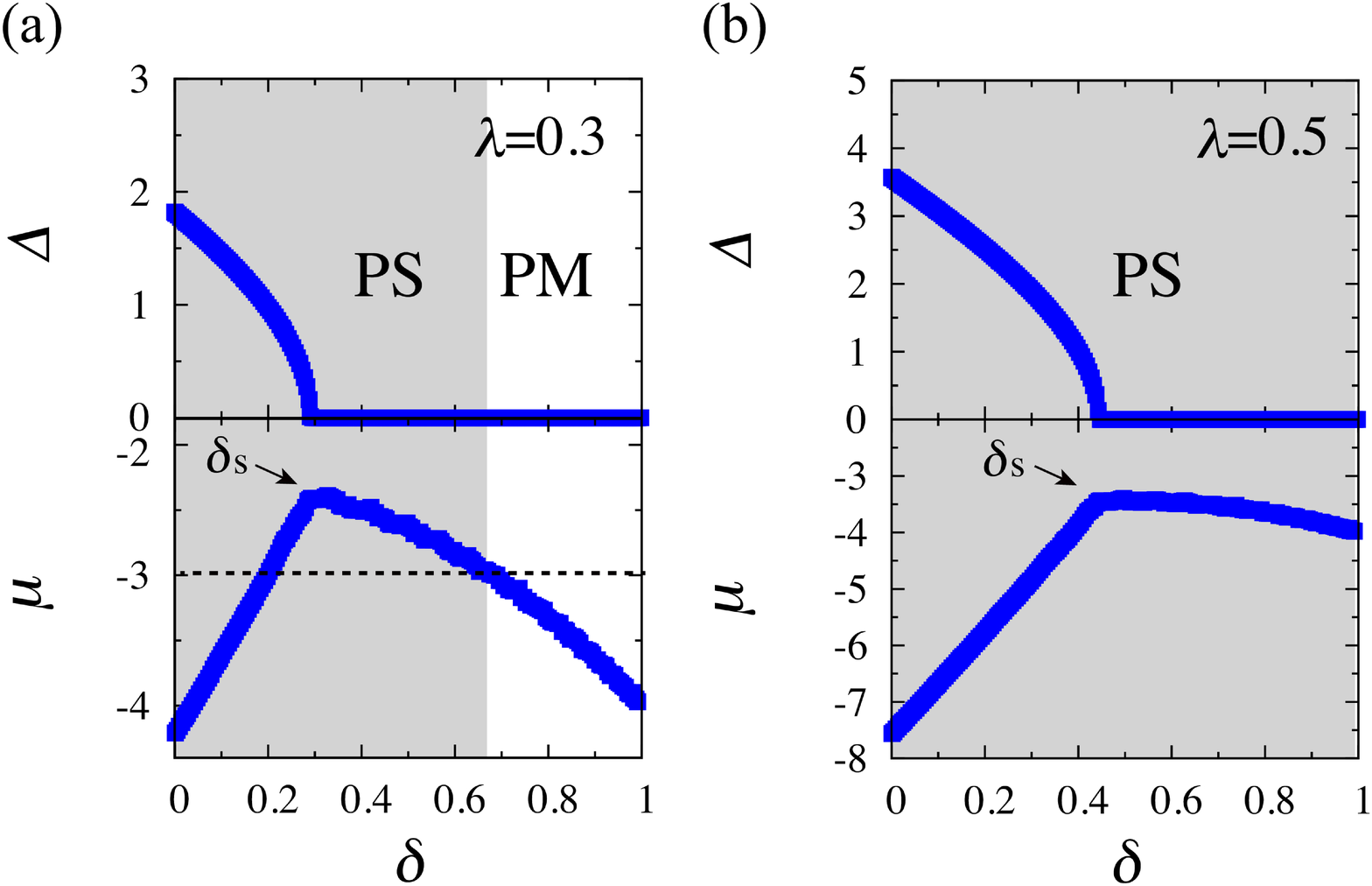}
\caption{(Color online) $\delta$-dependence of $\Delta$ and $\mu$ for (a) $\lambda=0.3$ and (b) $\lambda=0.5$. The system size is $40 \times 40$. Dashed lines in (a) represents the line for the Maxwell constructions. }\label{fig:phy_adiabatic}\end{center}
\end{figure}

\section{Energy per hole in doped regions}
In the main text, we have shown the chemical potential $\mu (\delta)$ to identify the PS region. Alternative way which is often used is to see the energy per hole defined by
\begin{eqnarray}
  e_{h} (\delta) = (E/N - E_0/N)/\delta.
\end{eqnarray}
Here, $E_0$ is the energy at $\delta$=0. If $e_h(\delta)$ has a minimum at $\delta=\delta_c$, a PS region is identified as $\delta < \delta_c$. In Fig. {\ref{fig:suppl_eh}, we plot this quantity for two parameter sets which are adopted in the main text. Minima are clearly seen and the PS regions are perfectly consistent with Fig. 5.

\begin{figure}[H]
\begin{center}
  \includegraphics[width=7cm]{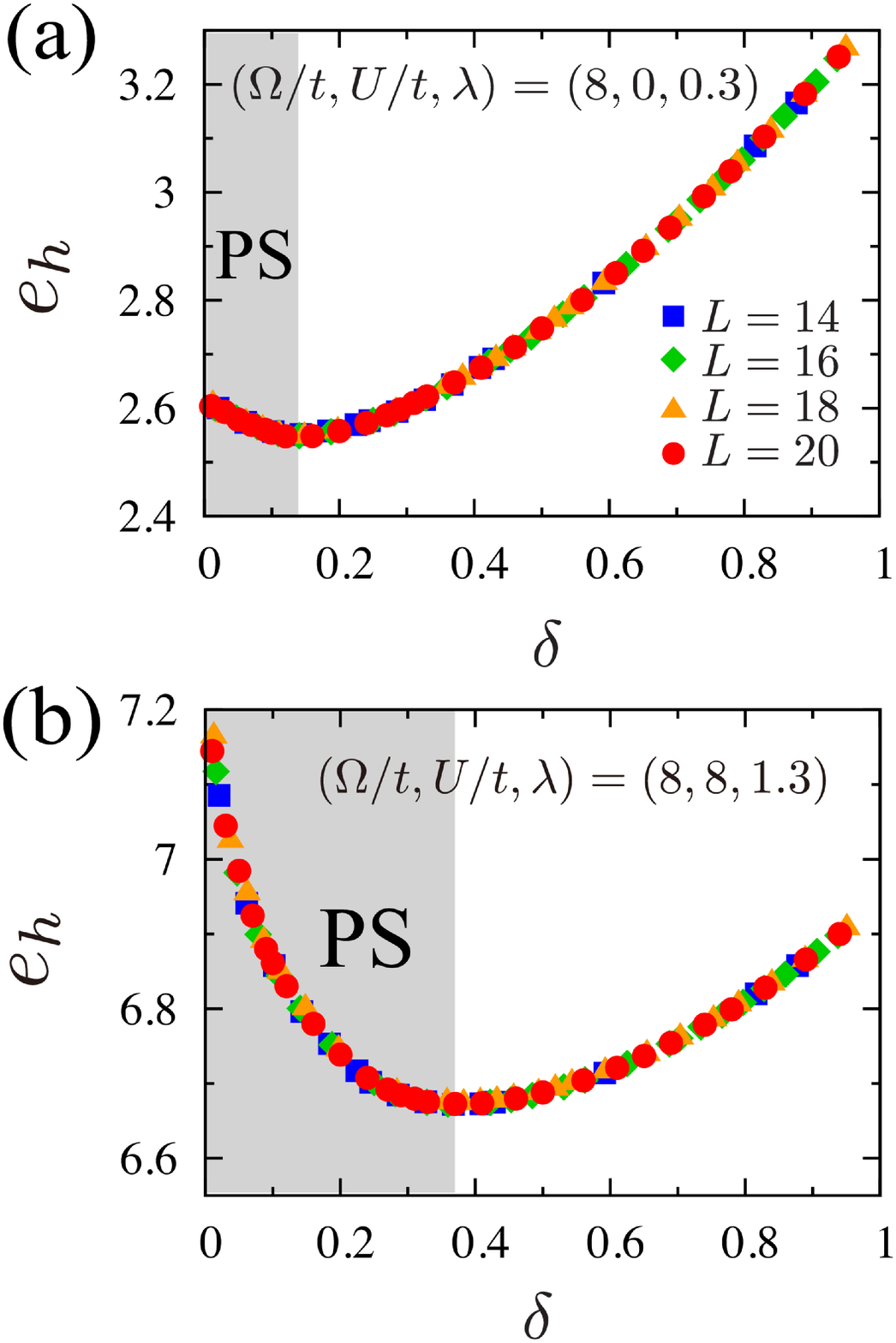}
\caption{(Color online) $\delta$-dependence of $e_h$ for (a) ($\Omega/t, U/t, \lambda$) = (8, 0, 0.3) and (b) ($\Omega/t, U/t, \lambda$) = (8, 8, 1.3).}\label{fig:suppl_eh}\end{center}
\end{figure}

\section{Possibility of incommensurate orders}
In the present study, we assumed a $2 \times 2$ sublattice structure for $\{f_{ij}\}$, and thus disregarded the possibility of incommensurate orders which may appear at finite $\delta$\cite{vekic1992}. The inclusion of such states may modify our phase diagram. Here, we discuss its possibility with additional data. 

In Fig. {\ref{fig:phy_incomme}, we show results for $(\Omega/t, U/t, \lambda)$ = (1, 0, 0.25), which include a charge-density-wave (CDW) state with long periodicity ${\bm l}=(l_x, l_y)$. Here, $l_{\alpha}$ represents the period in the $\alpha$-direction and we consider ${\bm l}=(8, 2)$, (10, 2), and (12, 2) which have peaks at ${\bm q}=(\pi-2\pi/l_x, \pi)$ in $S_c({\bm q})$. To obtain these states, we extended the sublattice structure to $8 \times 2$, $10 \times 2$, and $12 \times 2$, respectively. In addition to them, we show a CDW state with ${\bm l}=(2, 2)$ which is considered in the main text. As seen in Fig. {\ref{fig:phy_incomme}, we found that CDW states with ${\bm l}=(10, 2)$ or (12, 2) have lower energies than the states with ${\bm l}=(2, 2)$ in some $\delta$s, whereas the states with ${\bm l}=(8, 2)$ have higher energies. This suggests the presence of incommensurate orders in the thermodynamic limit. However, the inclusion of these states do not change the result of the Maxwell construction (the dashed line) and they are unstable against the phase separation.

\begin{figure}[H]
\begin{center}
  \includegraphics[width=6.5cm]{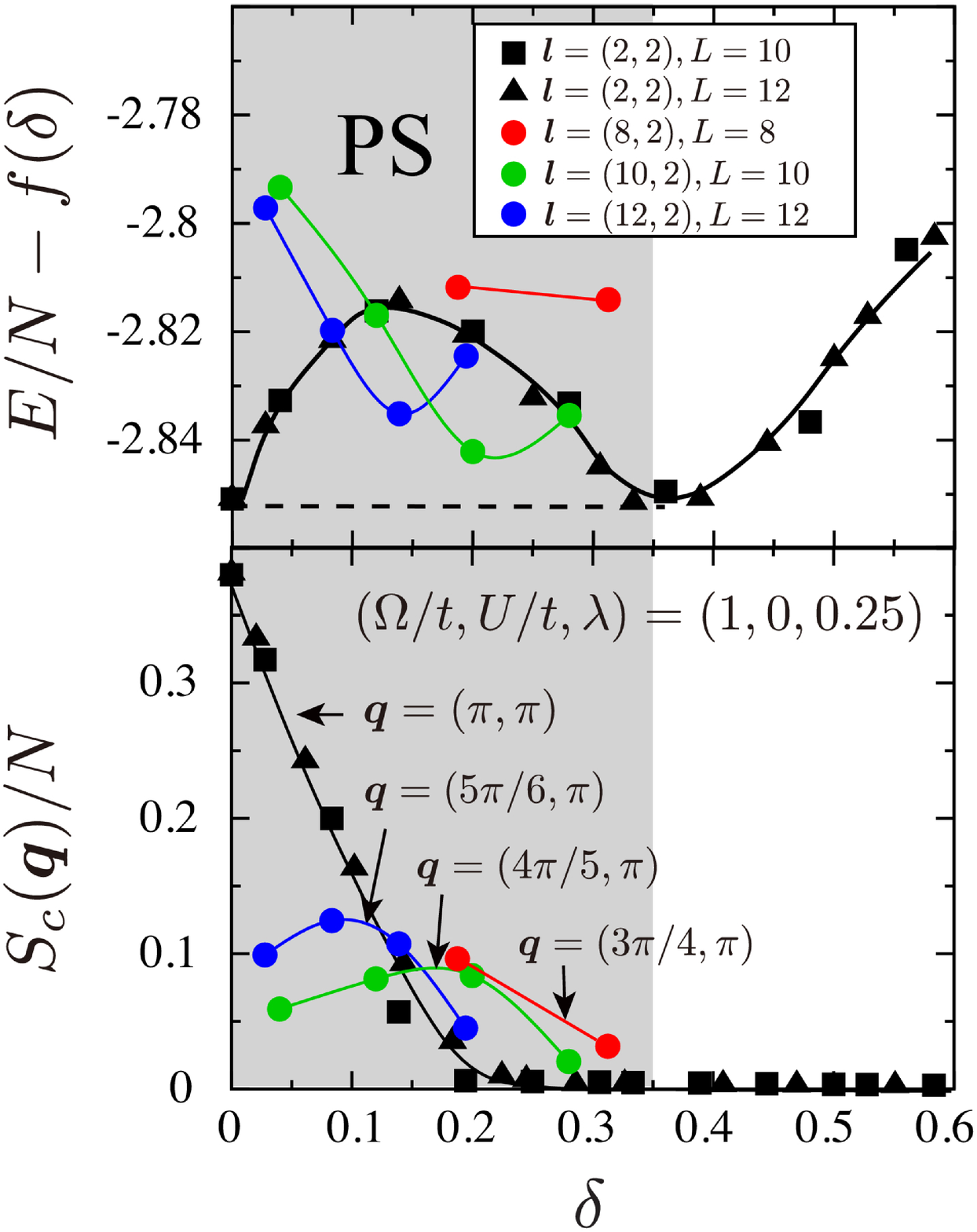}
\caption{(Color online) $E/N$ and $S_c({\bm q})/N$ for CDW states with ${\bm l}=(l_x, l_y)$. The parameter set is $(\Omega/t, U/t, \lambda) = (1, 0, 0.25)$. For ${\bm l}=(2, 2)$, the ground states under the $2\times 2$ sublattice structures are shown. An unimportant linear term $f(\delta) \propto \delta$ is subracted from $E/N$. The Maxwell construction is performed as denoted by the dashed line.}\label{fig:phy_incomme}\end{center}
\end{figure}

\end{document}